\documentclass[10pt,conference]{IEEEtran}
\usepackage[T1]{fontenc}
\usepackage{xcolor}
\usepackage{graphicx}
\usepackage{relsize}
\usepackage{enumitem}
\usepackage{color}
\usepackage{colortbl}
\usepackage{amssymb}
\usepackage{tabularx}
\usepackage{pifont}
\usepackage{soul}

\newcommand{\cmark}{\ding{51}} 
\newcommand{\dgc}{0.8}
\newcommand{\lgc}{0.9}
\definecolor{dg}{rgb}{\dgc, \dgc, \dgc}
\definecolor{lg}{rgb}{\lgc, \lgc, \lgc}

\usepackage{float}
\usepackage{multirow} 

\usepackage{hyperref}
\hypersetup{
colorlinks   = true,
citecolor    = blue,
urlcolor    =  blue
}
\usepackage{amsmath} 

\usepackage{booktabs}
\usepackage{makecell}
\usepackage{ragged2e}
\usepackage{xcolor}
\usepackage{tabularx}
\newcolumntype{L}{X}
\newcolumntype{C}{>{\centering\arraybackslash}X}
\newcolumntype{R}{>{\raggedleft\arraybackslash}X}
\newcommand{\footURL}[1]{\footnote{\url{#1}}}

\usepackage[numbers,square,sort&compress]{natbib}
\usepackage{todonotes}

\begin{document}

\title{TripleBound: Triplet-Guided Heterogeneous Graph Learning for Microservice Decomposition}


\author{
\IEEEauthorblockN{
Mineth Weerasinghe\IEEEauthorrefmark{1},
Himindu Kularathne\IEEEauthorrefmark{1},
Methmini Madhushika\IEEEauthorrefmark{1},
Danuka Lakshan\IEEEauthorrefmark{1}\\
Nisansa de Silva\IEEEauthorrefmark{1},
Adeesha Wijayasiri\IEEEauthorrefmark{1},
Srinath Perera\IEEEauthorrefmark{2}
}
\IEEEauthorblockA{
\IEEEauthorrefmark{1}Department of Computer Science and Engineering, University of Moratuwa, Sri Lanka\\
Email: \{mineth.21, himindu.21, methmini.21, dhanuka.21, NisansaDdS, adeeshaw\}@cse.mrt.ac.lk
}
\IEEEauthorblockA{
\IEEEauthorrefmark{2}WSO2 LLC\\
Email: srinath@wso2.com
}
}

\maketitle

\begin{abstract}
Cloud computing and DevOps have made microservices a common architecture for scalable, maintainable software systems. However, migrating monoliths to microservices remains challenging due to tight coupling and unclear service boundaries. Existing decomposition approaches typically rely on either structural dependencies or semantic similarity signals, but rarely integrate both within a unified representation learning objective.
This paper proposes TripleBound, a hybrid framework for automated monolith-to-microservices decomposition that augments a heterogeneous graph neural network with weakly supervised triplet constraints derived from parser-inferred service groups based on package structure, naming conventions, and code location. Rather than combining independently trained structural and organizational representations after training, TripleBound injects triplet-based constraints directly into the shared structural latent space, enabling both signals to be jointly optimized during representation learning.
Structural dependencies are captured using CHGNN, which models the monolith as a heterogeneous graph with program nodes, resource nodes, CALL edges, and CRUD edges. Semantic relationships are incorporated through triplet constraints generated from service groups inferred by a parser using cues such as package structure, naming conventions, and code location. 
Evaluation on AcmeAir, DayTrader, PlantsByWebSphere, and JPetStore shows that TripleBound achieves the highest composite decomposition score under the selected weighting on AcmeAir, DayTrader, and JPetStore compared to CHGNN and MonoEmbed, while CHGNN remains stronger on PlantsByWebSphere. Per-metric analysis reveals trade-offs: gains in structural modularity and inter-partition coupling are accompanied by higher entity distribution imbalance on some datasets. Alternative composite weightings preserve TripleBound's first-place ranking on AcmeAir and DayTrader but not on JPetStore, showing that the aggregate ranking is metric-dependent.
\end{abstract}


\section{Introduction}

Enterprise systems often begin as monoliths because they are simple to build, but become difficult to maintain, scale, and extend as they grow. The widespread adoption of cloud computing~\cite{armbrust2010cloud} and DevOps practices~\cite{kim2016devops, balalaie2016devops} has further accelerated demand for modular, independently deployable architectures. Microservices address these challenges by decomposing applications into small, focused services~\cite{ newman2021building, dragoni2017microservices,weerasinghe2026monolith}. Systematic mapping studies show that microservice research has grown around cloud deployment, architecture recovery, service identification, and migration support~\cite{pahl2016microservices, difrancesco2019architecting}. Despite its advantages, migrating monoliths to microservices remains complex~\cite{soldani2018pains, taibi2017processes, difrancesco2018industrial} and requires careful identification of service boundaries and granularity~\cite{gysel2016servicecutter, auer2021monolithic, hassan2020granularity, mohottige2025reengineering}.

Current decomposition techniques mainly rely on either structural dependencies or semantic analysis of source code. Structural approaches ~\cite{aldebagy2021microservice, mitchell2006automatic, saidani2019automated, mazlami2017microservice} analyze call graphs and dependency networks to capture interactions between software components. Semantic approaches use embeddings derived from source code to capture contextual similarities between components.
However, approaches that rely on a single signal often fail to capture the full characteristics of software systems. Structural methods may ignore semantic relationships between components, while semantic methods may overlook critical dependency structures.

To address these limitations, this paper proposes TripleBound, a hybrid decomposition framework that integrates structural and source-level organizational signals through joint representation learning. The structural view is captured using a heterogeneous graph neural network~\cite{mathai2021chgnn}, which represents the monolithic system as a heterogeneous graph capable of modeling structural and behavioral relationships~\cite{mancoridis1998automatic} among software components. The organizational view is incorporated through weakly supervised triplet constraints derived from inferred service groups, encouraging related code elements to cluster more closely within the shared structural representation space. By optimizing these complementary signals within a unified training objective, TripleBound aims to generate cohesive and architecturally consistent microservices.

The evaluation uses four open-source Java monoliths frequently examined in decomposition and modernization research~\cite{kalia2021mono2micro,sellami2022hierarchical,abgaz2023decomposition,saied2024migration}. AcmeAir represents a REST-based airline reservation system, DayTrader is a comparatively larger Java EE transaction-processing application, PlantsByWebSphere models an online nursery, and JPetStore is a compact MyBatis-based commerce application. Together, they provide variation in application size, framework, dependency structure, and source-code organization; their detailed characteristics are reported in the experimental evaluation.
The main contributions of this paper are:

\begin{enumerate}[label=(\arabic*)]
    \item A hybrid monolith-to-microservices decomposition framework that jointly integrates heterogeneous graph-based structural learning and weakly supervised triplet-based constraint learning within a unified representation learning process.

    \item We introduce a parser-inferred triplet-guided graph embedding strategy in which weakly supervised triplet constraints are injected directly into the structural embedding space, rather than combining structural outputs with independently trained representations.

    \item We formulate a joint optimization objective that combines node reconstruction, edge reconstruction, clustering consistency, semantic triplet loss, and communication-aware coupling reduction to learn microservice embeddings.

    \item We evaluate TripleBound on four benchmark monoliths and compare it against available structural and semantic baselines using Structural Modularity (SM), Interface Number (IFN), Inter-Partition Communication (ICP), Non-Extreme Distribution (NED), and a composite score.
\end{enumerate}

\section{Related Work}

\begin{table}[t]
\centering
\caption{Comparison of related approaches}
\label{tab:related_work}
\footnotesize
\setlength{\tabcolsep}{6pt}
\renewcommand{\arraystretch}{1.15}

\begin{tabularx}{0.48\textwidth}{lccX}
\toprule
\textbf{Approach} 
& \textbf{Struct.} 
& \textbf{Sem.} 
& \textbf{Limitation} \\
\midrule

CHGNN~\cite{mathai2021chgnn}
& \cmark
& --
& Does not explicitly model semantic similarity. \\

MonoEmbed~\cite{sellami2025contrastive}
& Limited
& \cmark
& Limited awareness of architectural dependencies. \\

Mono2Micro~\cite{kalia2021mono2micro}
& \cmark
& Partial
& Uses multiple signals, but not through joint representation learning. \\

CoGCN~\cite{desai2021gnn}
& \cmark
& --
& Mainly focuses on graph structure and ignores semantic constraints. \\

\textbf{TripleBound}
& \cmark
& \cmark
& Depends on the quality of inferred triplets. \\
\bottomrule

\end{tabularx}
\end{table}
Automated monolith-to-microservices decomposition has been studied using structural, semantic, graph-based, and hybrid techniques ~\cite{sellami2025contrastive,abgaz2023decomposition,fosci2020,kalia2020mono2micro,sellami2022hierarchical}. Structural approaches analyze dependencies such as method calls, data access, and runtime interactions to identify highly connected components~\cite{levcovitz2016technique, mazlami2017microservice, li2019dataflow, filippone2023graphcluster}. These methods are useful for preserving technical dependencies, but they may ignore domain-level semantic relationships between code elements.

Semantic approaches use identifiers, comments, package names, or code embeddings~\cite{allamanis2018survey} to capture functional similarity between components~\cite{aldebagy2021microservice, al2021semantic, trabelsi2022legacy}. For example,~\citet{brito2021identification} apply topic modelling to lexical information extracted from source code, where the inferred topics correspond to domain terms and are used to identify candidate microservices. Methods such as MonoEmbed~\cite{sellami2025contrastive} leverage pre-trained transformer-based language models~\cite{vaswani2017attention} and code-specific embeddings~\cite{feng2020codebert} enhanced through contrastive learning~\cite{chen2020simclr} and LoRA fine-tuning~\cite{hu2021lora} to group semantically related code elements. However, semantic similarity alone may overlook important architectural constraints such as inter-service communication and database coupling.

Graph-based approaches are widely used in microservice reengineering because
software systems can be represented as graphs in which classes, methods,
components, database tables, or other system entities are modeled as vertices,
while dependencies, method calls, and coupling relationships are modeled as
edges~\cite{mohottige2025reengineering}. Recent extraction methods have also explored knowledge graphs with constrained community detection and graph deep clustering with multiple software views~\cite{li2022knowledgegraph, qian2023gdc}. Following this direction, CHGNN
models the monolith as a heterogeneous graph containing program nodes, resource
nodes, CALL edges, and CRUD edges. This enables the model to learn rich
structural representations of software components~\cite{mathai2021chgnn, desai2021graph, lecrivain2025mono2rest}. However, CHGNN mainly
focuses on structural dependency learning and does not explicitly optimize semantic similarity during training.

Hybrid approaches ~\cite{hierdecomp2019, sellami2022hierarchical, krause2020static} combine multiple signals such as structural dependencies, runtime traces, database access, and semantic information. Recent deep-learning approaches similarly construct structural and semantic views before clustering learned representations~\cite{qian2023gdc, wei2025gcvcg}. Tools such as Mono2Micro~\cite{kalia2021mono2micro} and DEEPLY~\cite{yedida2022deeply} use several analysis signals or objective-tuning strategies, but typically combine them through separate stages rather than joint optimization. In contrast, TripleBound integrates structural graph learning and weakly supervised triplet-based constraint learning within a single unified training objective.

Several empirical and survey studies have contributed comparative analyses of decomposition approaches. \citet{taibi2021empirical} conducted an empirical comparison of multiple microservice identification techniques applied to a shared monolithic system, revealing that different approaches can produce substantially different decompositions on the same codebase even when targeting the same number of services. This highlights the importance of multi-metric evaluation when comparing methods. A systematic review by~\citet{abgaz2023decomposition} catalogued existing decomposition approaches by input type, algorithmic technique, and evaluation methodology, identifying the absence of standardized benchmarks as a key barrier to rigorous cross-study comparison.
The study by~\citet{weerasinghe2026monolith} is a comprehensive comparison of state-of-the-art microservice decomposition methods against benchmark data sets established in the domain.  

Additional tool-based and learning-driven approaches have been proposed to address specific aspects of the decomposition problem. CARGO~\cite{nitin2022cargo} applies AI-guided dependency analysis to identify tightly coupled component clusters, leveraging graph-based structural reasoning and constraint propagation to support migration planning decisions. \citet{huang2024abmsc} introduce an attention-based bidirectional microservice clustering framework that incorporates directional relationship weighting to better capture asymmetric dependencies between software components. FoSCI~\cite{fosci2020} proposes a feature-oriented service clustering and identification approach that groups code elements according to feature-level analysis, providing an alternative decomposition perspective that complements purely structural or semantic strategies. RapidMS~\cite{zhang2023rapidms} takes a complementary direction by generating initial microservice structures from requirements models, demonstrating that decomposition guidance can also be derived from high-level design artifacts rather than source code analysis alone. Mo2oM~\cite{ziabakhsh2025mo2om} formulates microservice extraction as a soft clustering problem, combining deep semantic code embeddings with structural dependency graphs to allow components to belong probabilistically to multiple services, thereby relaxing the hard partition assumption that most decomposition methods, including the approach proposed in this paper, rely on.

Another important distinction among existing decomposition approaches is the type of input required. Some approaches depend on manual or semi-manual inputs such as domain models~\cite{evans2004ddd}, use cases, data-flow diagrams, design artifacts, or user-provided system knowledge ~\cite{gysel2016servicecutter, li2019dataflow, li2022knowledgegraph}. While such inputs can improve domain awareness, they increase the effort required from developers and domain experts and may limit automation. In contrast, code-driven approaches aim to reduce this manual effort by extracting structural and semantic information directly from the monolithic application.

Combining structural and semantic signals consistently yields more coherent service boundaries than either alone~\cite{hierdecomp2019, sellami2022hierarchical}, motivating the joint optimization objective proposed in this paper.

\section{TripleBound Approach}

\subsection{Framework Overview}

\begin{figure}[t]
\centering
\includegraphics[
    width=1.00\linewidth,
]{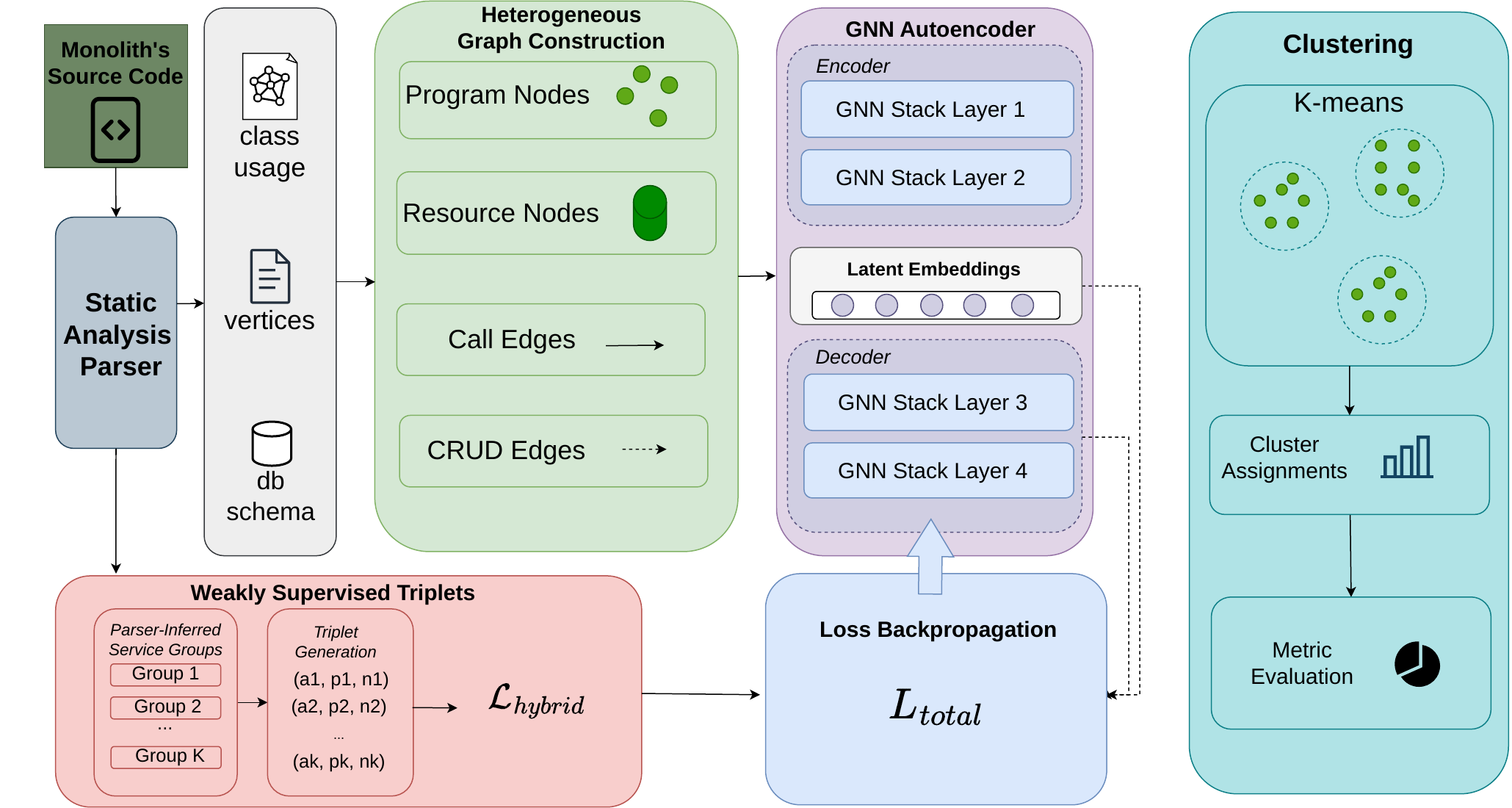}
\caption{Architecture of TripleBound}
\label{fig:architecture}
\end{figure}

TripleBound integrates structural and semantic analysis to generate microservice decompositions. Figure~\ref{fig:architecture} illustrates the overall architecture of TripleBound.
The parser constructs a heterogeneous graph containing program and resource nodes connected by CALL and CRUD edges, while also inferring service groups from package structure, naming conventions, and code location. Type-specific transformations, Node2CommonSpace and Edge2CommonSpace, project node and edge features into a shared representation space processed by an edge-aware graph autoencoder~\cite{kipf2016variational}. The autoencoder consists of stacked GNN encoder layers, a bottleneck representation, and a decoder responsible for reconstructing node and edge information while preserving the structural properties of the original application.

Triplets generated from the parser-inferred groups are applied to the same encoder to guide the learned latent representations. This joint learning process encourages related components to remain close while separating components belonging to different inferred service groups. The resulting embeddings are partitioned using the K-means clustering algorithm~\cite{lloyd1982kmeans}, following the clustering approach used by CHGNN due to its simplicity and effectiveness in partitioning embedding spaces. The number of clusters, $K$, is set to the number of inferred service groups. For the benchmark systems in this study, the resulting $K$ values range from 4 to 6, as shown in Table~\ref{tab:dataset_stats}. The resulting clusters represent candidate microservices and are evaluated using SM, IFN, ICP, and NED, as shown in the output stage of Figure~\ref{fig:architecture}.

\subsection{Heterogeneous Graph Construction}

To capture the structural characteristics of the monolithic system, we model the codebase as a heterogeneous graph following the representation used in CHGNN~\cite{mathai2021chgnn}, building on heterogeneous relational graph modeling principles~\cite{schlichtkrull2018rgcn}.

The graph consists of two primary types of nodes:
\begin{itemize}
    \item \textbf{Program nodes}, representing code elements such as classes and services
    \item \textbf{Resource nodes}, representing external entities such as database tables~\cite{mathai2021chgnn}
\end{itemize}

Node relationships are modeled using typed edges:
\begin{itemize}
    \item \textbf{CALL edges}, capturing invocation relationships between program components
    \item \textbf{CRUD edges}, representing interactions between program components and database resources
\end{itemize}


Each node and edge is associated with a feature representation encoding its structural properties. Because program nodes, resource nodes, CALL edges, and CRUD edges may have different feature formats, separate type-specific transformations are used to map them into a common latent space before they are processed by the graph encoder.
This representation allows the model to jointly capture intra-code dependencies and data access patterns, which are critical for identifying cohesive microservice boundaries.

While the graph construction follows the CHGNN formulation, it serves as the structural backbone of TripleBound, enabling seamless integration with the semantic triplet learning objective. In particular, the unified latent space learned from this graph facilitates the incorporation of both structural dependencies and semantic similarity constraints during training.

\subsection{Structural Representation Learning}

In analyzing the system, we leverage the fact that most programming languages and frameworks used for building web services follow layered architectural patterns~\cite{richards2015software, trabelsi2022legacy, zaragoza2022layered}. In such architectures, methods in the presentation layer expose public interfaces and handle incoming requests, while lower-layer methods are invoked to process business logic and interact with the persistence layer and database. These layered interactions naturally form structural dependency patterns within the monolithic system.

Following~\citet{weerasinghe2026monolith}, we adopt the heterogeneous graph neural network (CHGNN) framework~\cite{mathai2021chgnn}, which models the system as an edge-aware graph autoencoder.

Given the heterogeneous graph constructed in the previous section, the model learns low-dimensional embeddings for each node by encoding both node attributes and typed edge relationships. Due to the presence of multiple node and edge types, type-specific transformations are first applied to project features into a shared latent space.

The encoder consists of stacked graph neural network layers~\cite{hamilton2017graphsage, wu2021gnn} that propagate and aggregate information across neighboring nodes, taking into account both structural connectivity and edge semantics. This process produces a latent embedding $h_i$ for each node $i$, capturing its structural role within the system.

A decoder is then used to reconstruct node attributes and typed edge features.
The reconstruction objective ensures that the learned embeddings preserve important structural dependencies, such as method invocations and data access patterns.
These structural embeddings serve as the foundation for the subsequent semantic triplet learning stage.

\subsection{Semantic Triplet Construction}

To incorporate semantic relationships between software components, we employ a triplet-based metric learning strategy following the general triplet-network formulation~\cite{hoffer2015deep} and inspired by \textit{MonoEmbed}~\cite{sellami2025contrastive}, as these showed high promise in the comparative study conducted by~\citet{weerasinghe2026monolith}.

Given the absence of explicit ground-truth microservice boundaries in real-world monolithic systems, we treat parser-inferred service groups as weak supervisory signals, following the broader idea of weakly supervised contrastive learning, where inferred similarity relations are used to guide representation learning~\cite{zheng2021weakly}. 
Specifically, the parser produces a service-to-node mapping, where each group contains code elements that are logically associated based on factors such as package structure, naming conventions, and location within the codebase.

Formally, let $\mathcal{S} = \{S_1, S_2, ..., S_k\}$ denote the set of inferred service groups, where each $S_i$ is a set of nodes (classes). Using this grouping, triplets are constructed as follows:

\begin{itemize}
    \item \textbf{Anchor ($a$)}: a randomly selected node from a service $S_i$
    \item \textbf{Positive ($p$)}: another node from the same service $S_i$
    \item \textbf{Negative ($n$)}: a node sampled from a different service $S_j$, where $i \neq j$
\end{itemize}
Since our service groups are parser-inferred weak labels rather than manually verified ground truth, we avoid hard-negative mining and instead randomly sample negatives from different inferred service groups. This conservative strategy reduces the risk of unstable optimization caused by hard-negative triplets, which have been shown to produce bad local minima under standard triplet loss training~\cite{xuan2020hard}.
The triplets are not used to train a separate semantic embedding model. Instead, the anchor, positive, and negative nodes are passed through the same encoder used for structural graph representation learning. Therefore, the triplet loss directly shapes the CHGNN latent space. This design differs from approaches that compute structural and semantic representations independently and combine them only after training. By applying triplet constraints to the shared latent embeddings, the model learns a unified representation in which structurally dependent components and semantically related components are jointly organized.
Certain nodes may not be associated with any inferred service group because the parser cannot classify them using package structure, naming conventions, or code location. Such nodes, including shared utility and infrastructure classes, are excluded from triplet generation but remain in the heterogeneous graph for structural representation learning.

\subsection{Hybrid Training Objective}

\begin{figure*}[!t]
\centering
\includegraphics[
width=0.84\textwidth
]{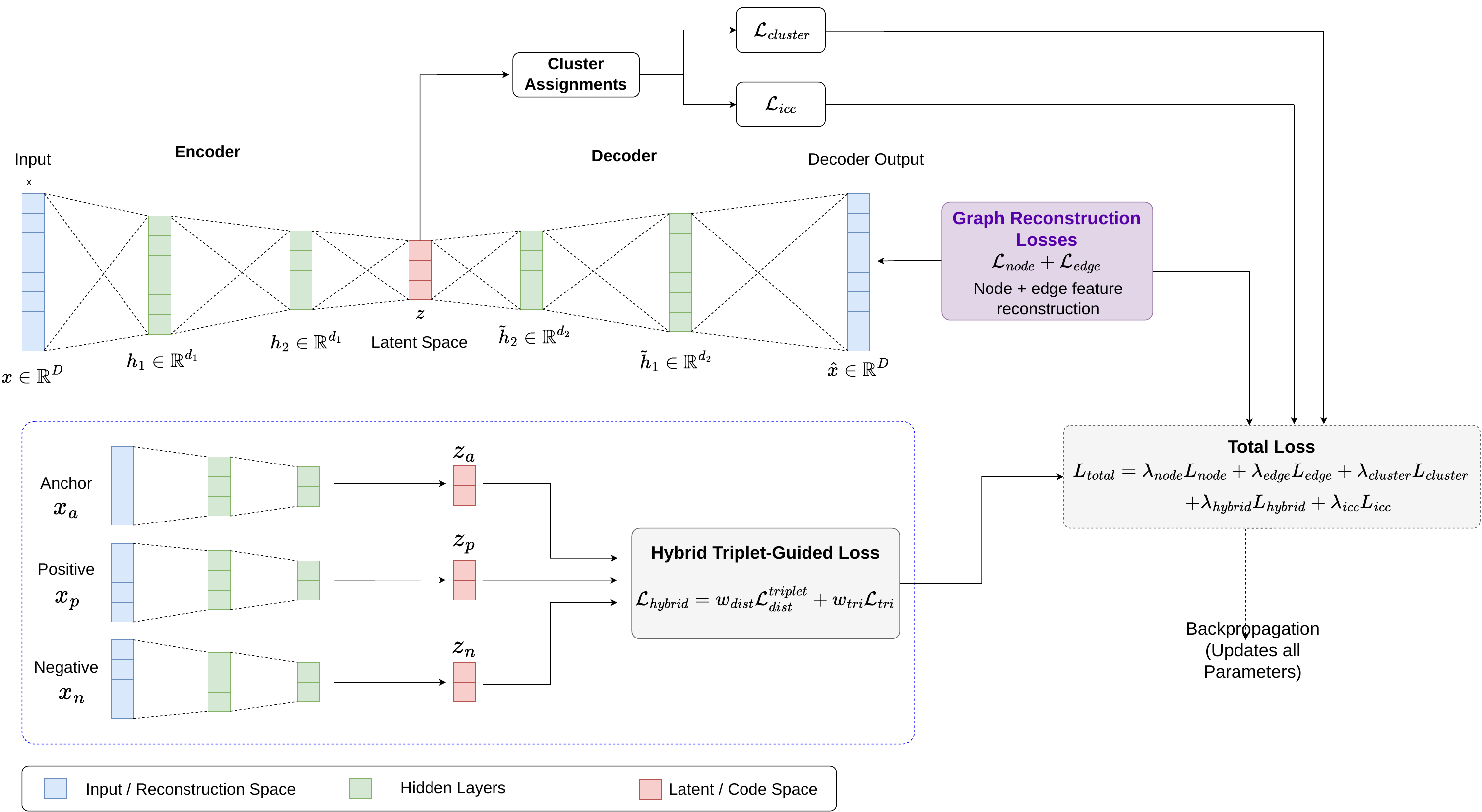}
\caption{Internal training architecture of TripleBound.}
\label{fig:internal_architecture}
\end{figure*}

Structural dependencies alone are often insufficient to capture domain-level relationships between software components, while purely semantic approaches may overlook critical architectural constraints. To address this limitation, we propose a hybrid training objective that jointly optimizes structural and semantic representations.

Most graph neural networks follow a message-passing mechanism, where the vector representation of a node is updated by combining its own features with aggregated information from its neighboring nodes~\cite{desai2021graph}. TripleBound combines node and edge reconstruction losses with a hybrid triplet-guided loss, clustering consistency, and communication-aware regularization. This enables the model to preserve both dependency structures and semantic similarity relationships within a unified embedding space.

\subsubsection{Joint Loss Formulation}

To integrate weak semantic supervision into the shared structural embedding space, we define a hybrid loss component. This component combines a triplet-local distance regularizer with a semantic triplet margin loss. The distance regularizer preserves the encoded/decoded representations of the nodes participating in each sampled triplet, while the triplet margin loss encourages nodes from the same parser-inferred service group to be closer than nodes from different groups.

Let \(\mathcal{T}\) denote the set of sampled triplets. Each triplet \((a,p,n) \in \mathcal{T}\) consists of an anchor node \(a\), a positive node \(p\) sampled from the same parser-inferred service group, and a negative node \(n\) sampled from a different group. Let \(x_{e,i}\) and \(x_{d,i}\) denote the encoded and decoded representations of node \(i\), respectively, and let \(h_i\) denote the latent embedding used for triplet learning.

The triplet-local distance regularization term is defined as:

\begin{equation}
L_{dist}^{triplet}
=
\frac{1}{|\mathcal{T}|}
\sum_{(a,p,n)\in\mathcal{T}}
\frac{1}{3}
\sum_{i\in\{a,p,n\}}
\|x_{d,i}-x_{e,i}\|_3^3
\end{equation}

This term is computed only over the anchor, positive, and negative nodes in each sampled triplet. The semantic triplet loss is defined as:

\begin{equation}
L_{tri}
=
\frac{1}{|\mathcal{T}|}
\sum_{(a,p,n)\in\mathcal{T}}
\max
\left(
0,
\|h_a-h_p\|_2^2
-
\|h_a-h_n\|_2^2
+
\beta
\right)
\end{equation}

where \(\beta\) is the triplet margin.

To dynamically balance the two objectives during training, we define the distance and triplet weights as:

\begin{equation}
w_{dist} =
\max\left(\alpha, \frac{N-e}{N}\right)
\end{equation}

\begin{equation}
w_{tri} =
\min \left(1 - \alpha, \frac{e}{N} \right)
\end{equation}

where $0 < \alpha < 1$ is the hybrid-balance hyperparameter, $e$ is the current training epoch and $N$ is the total number of epochs. The hybrid loss component is then defined as:

\begin{equation}
\mathcal{L}_{hybrid}
=
w_{dist}\mathcal{L}_{dist}^{triplet}
+
w_{tri}\mathcal{L}_{tri}
\end{equation}

The first term corresponds to triplet-local distance regularization, while the second term corresponds to semantic triplet supervision. Thus, \(\mathcal{L}_{hybrid}\) regularizes the representations of triplet nodes while simultaneously enforcing parser-inferred grouping constraints in the latent space.

The weighting mechanism dynamically balances local distance preservation and semantic triplet supervision during training. In early epochs, the model gives higher emphasis to the triplet-local distance term, while in later epochs the triplet margin loss receives greater relative emphasis.
In our experiments, we set the hybrid-balance parameter to $\alpha = 0.7$, which places greater emphasis on semantic consistency during later training stages while still preserving structural information in early epochs.
The final loss used during training is implemented as a weighted combination of multiple components:

\begin{equation}
\begin{aligned}
\mathcal{L}_{total} =
&\lambda_{node}\mathcal{L}_{node}
+\lambda_{edge}\mathcal{L}_{edge}
+\lambda_{cluster}\mathcal{L}_{cluster} \\
&+\lambda_{hybrid}\mathcal{L}_{hybrid}
+\lambda_{icc}\mathcal{L}_{icc}
\end{aligned}
\end{equation}

where \(\mathcal{L}_{node}\) and \(\mathcal{L}_{edge}\) are the graph-wide node and edge reconstruction losses, \(\mathcal{L}_{cluster}\) enforces clustering consistency, \(\mathcal{L}_{hybrid}\) is the hybrid loss component, and \(\mathcal{L}_{icc}\) is the proposed Inter-Cluster Communication Loss.  The coefficients \(\lambda_{node}\), \(\lambda_{edge}\), \(\lambda_{cluster}\), \(\lambda_{hybrid}\), and \(\lambda_{icc}\) are configurable hyperparameters set before training to control the relative contribution of each objective. This formulation allows the model to jointly optimize structural, semantic, and communication-aware objectives while keeping reconstruction independent of the triplet sampling process.

To reduce inter-service communication during training, we introduce an
Inter-Cluster Communication Loss, denoted as \(\mathcal{L}_{icc}\). Before
defining this term, we define the clustering objective used to encourage compact
service partitions. Given node embeddings \(h_i\), cluster centroids \(C_k\),
and a binary assignment matrix \(M\), the clustering loss is defined as:

\[
\mathcal{L}_{cluster}
=
\sum_{i \in V}
\sum_{k=1}^{K}
M_{i,k}
\left\|
h_i - C_k
\right\|_2^2
\]

where \(M_{i,k}=1\) if node \(i\) is assigned to cluster \(k\), and
\(M_{i,k}=0\) otherwise. Minimizing \(\mathcal{L}_{cluster}\) pulls node
embeddings closer to their assigned cluster centroids, improving cluster
compactness.

Since the final ICP metric is computed after hard cluster assignment and is not
directly differentiable, we define a soft communication-aware proxy based on
cluster assignment probabilities. Given node embeddings and cluster centroids, we
compute the distance between each node and each centroid and apply a softmax over
the negative distances:

\[
p_i = \text{softmax}(-d(h_i, C))
\]

where \(p_i\) represents the probability distribution of node \(i\) over
clusters. For each dependency edge \((u,v)\), the probability that its
endpoints belong to the same cluster is computed as:

\[
P_{\text{same}}(u,v) = \sum_{k=1}^{K} p_{u,k} p_{v,k}
\]

The Inter-Cluster Communication Loss is then defined as:

\[
\mathcal{L}_{icc} =
\frac{1}{|E|}
\sum_{(u,v)\in E}
\left(1 - P_{\text{same}}(u,v)\right)
\]

Minimizing \(\mathcal{L}_{icc}\) encourages nodes connected by dependency edges
to be assigned to the same cluster, thereby reducing potential communication
across service boundaries. Although \(\mathcal{L}_{icc}\) is not the final ICP evaluation metric, it acts as a differentiable proxy that guides the
embedding space toward more communication-aware microservice boundaries.

During preliminary experiments, the original CHGNN structure loss caused
unstable optimization in the hybrid setting. Since it reconstructs adjacency
using raw pairwise dot products of node embeddings, it produced large gradients
when combined with semantic triplet and clustering objectives, dominating the
overall loss. This aligns with multi-task learning observations that objectives
with larger gradient magnitudes can hinder balanced training~\cite{chen2018gradnorm}.
Therefore, we removed the structure loss and retained node reconstruction, edge
reconstruction, clustering, semantic triplet, and communication-aware losses,
resulting in more stable training.

Figure~\ref{fig:internal_architecture} illustrates the internal training architecture: the encoder processes each component representation into a shared latent embedding; the decoder reconstructs the original input; and parser-inferred triplets are passed through the same encoder to apply semantic constraints, all optimized jointly via backpropagation.

\section{Experimental Evaluation}

\subsection{Experimental Setup}

All experiments were repeated for 30 runs with different random initializations, and the reported results correspond to the average values across these runs. The loss components were weighted as follows: \(\lambda_{node}=0.001\), \(\lambda_{edge}=0.1\), \(\lambda_{cluster}=0.7\), \(\lambda_{hybrid}=1.0\), and \(\lambda_{icc}=1.0\). These values differ slightly from the original CHGNN configuration and were empirically selected to better balance structural preservation and clustering quality in the hybrid setting.
A replication package is available at
\url{https://github.com/Mono2Distributed/triplet-guided-chgnn}.
The package includes the implementation, configuration files,
and scripts required to reproduce the reported experiments.
The JPetStore call graph generated for CHGNN compatibility is
publicly available on Hugging Face at
\url{https://huggingface.co/datasets/Min3th/jpetstore-chgnn-callgraph}.

\subsection{Datasets}

\begin{table*}[t]
\centering
\caption{Characteristics of benchmark systems}
\label{tab:dataset_stats}
\footnotesize
\setlength{\tabcolsep}{6pt}
\renewcommand{\arraystretch}{1.15}

\begin{tabular}{lrrrrrr}
\toprule
\textbf{Dataset} & 
\textbf{ICU Classes} & 
\textbf{Services/} & 
\textbf{DB/Resource} & 
\textbf{Inferred} & 
\textbf{Ignored} & 
\textbf{Clusters} \\
& & \textbf{Endpoints} & \textbf{Entries} & \textbf{Groups} & \textbf{Classes} & \\
\midrule

AcmeAir 
& 38 & 20 & 30 & 4 & 0 & 4 \\

DayTrader 
& 111 & 210 & 159 & 6 & 8 & 6 \\

PlantsByWebSphere  
& 36 & 47 & 40 & 5 & 0 & 5 \\

JPetStore
& 23 & 21 & 25 & 4 & 0 & 4 \\

\bottomrule
\end{tabular}
\end{table*}

We evaluate TripleBound using four open-source monolithic Java systems commonly used as evaluation subjects in microservice decomposition and modernization studies: 
\texttt{AcmeAir}\footURL{https://github.com/acmeair/acmeair}, \texttt{DayTrader}\footURL{https://github.com/WASdev/sample.daytrader7}, \texttt{PlantsByWebSphere}\footURL{https://github.com/WASdev/sample.plantsbywebsphere}, and \texttt{JPetStore}\footURL{https://github.com/mybatis/jpetstore-6}~\cite{kalia2021mono2micro,sellami2022hierarchical,abgaz2023decomposition,saied2024migration}. Other systems, such as Train Ticket~\cite{trainticket}, are also present in the literature but are designed as microservice-native applications rather than monolithic systems intended for decomposition, and are therefore not used in this evaluation.

AcmeAir is an airline reservation benchmark developed by IBM that represents a cloud-oriented web application with REST services, domain entities, and database/resource interactions.
DayTrader~\cite{daytrader2007} is a Java EE stock-trading benchmark that includes common enterprise components such as Servlets, JSPs, EJBs, JPA, JDBC, JMS, and transactions.
PlantsByWebSphere is a Java EE sample application for an online plant nursery, containing web and REST-based functionality with business entities, resource accesses, and service entry points.
JPetStore is a compact MyBatis-based Java web application containing account, catalog, cart, and order functionality.

Table~\ref{tab:dataset_stats} summarizes the dataset characteristics used in our experiments. ICU denotes Inter-Class Usage. Inferred Groups denotes the parser-inferred service groups used for triplet generation. Clusters denotes the final value of K, which is set to the number of inferred groups.

\subsection{Evaluation Metrics}

The quality of the generated decompositions is evaluated using metrics:

\subsubsection{Structural Modularity (SM)~\cite{fosci2020}}assesses the structural quality of a decomposition by considering both the cohesiveness of classes within each partition and the coupling between different partitions. A good decomposition should group classes that collaborate closely while reducing dependencies across service boundaries, which is consistent with the single responsibility principle. It is computed as shown in Equation~\ref{eq:sm}. In the equation, $scoh_i = \frac{c_{i,i}}{m_i^2}$ represents the cohesiveness within partition $i$. Similarly, $scop_{i,j} = \frac{c_{i,j}}{2(m_i*m_j)}$ represents the coupling between partitions $i$ and $j$. Higher SM values indicate better modular decomposition~\cite{mitchell2008}.

\begin{table*}[t]
\centering
\caption{Comparison of Decomposition Quality Across Datasets}
\label{tab:decomposition_quality}
\footnotesize
\setlength{\tabcolsep}{7pt}
\renewcommand{\arraystretch}{1.15}

\begin{tabular}{llccccc}
\toprule
\textbf{Dataset} & \textbf{Method} 
& \textbf{SM} $\uparrow$ 
& \textbf{IFN} $\downarrow$ 
& \textbf{ICP} $\downarrow$ 
& \textbf{NED} $\downarrow$ 
& \textbf{Composite} $\uparrow$ \\
\midrule

\multirow{3}{*}{DayTrader}
& CHGNN     & 0.13 & 5.70 & 0.55 & \textbf{0.50} & -0.3043 \\
& MonoEmbed & 0.13 & \textbf{1.35} & 0.59 & 0.66 & -0.4665 \\
& TripleBound   & \textbf{0.14} & 5.10 & \textbf{0.48} & 0.62 & \textbf{0.7708} \\
\midrule

\multirow{3}{*}{PlantsByWebSphere}
& CHGNN     & \textbf{0.17} & 3.60 & 0.51 & \textbf{0.20} & \textbf{0.5188} \\
& MonoEmbed & 0.11 & \textbf{1.55} & \textbf{0.26} & 0.82 & -0.4449 \\
& TripleBound   & 0.15 & 4.31 & 0.58 & 0.21 & -0.0738 \\
\midrule

\multirow{3}{*}{AcmeAir}
& CHGNN     & 0.11 & 2.50 & 0.37 & \textbf{0.00} & 0.2782 \\
& MonoEmbed & 0.01 & 3.06 & 0.47 & 0.56 & -1.1066 \\
& TripleBound   & \textbf{0.23} & \textbf{2.36} & \textbf{0.28} & 0.71 & \textbf{0.8284} \\
\midrule

\multirow{3}{*}{JPetStore}
& CHGNN  & 0.15 & 3.00 & 0.33 & \textbf{0.00} & 0.2046\\
& MonoEmbed & 0.03 & \textbf{2.10} & \textbf{0.29} & 0.72 & -0.4502\\
& TripleBound & \textbf{0.18} & 2.95 & 0.34 & 0.24 & \textbf{ 0.2455} \\
\bottomrule

\end{tabular}
\end{table*}

\setlength{\abovedisplayskip}{3pt}
\setlength{\belowdisplayskip}{3pt}
\begin{equation}
    SM = \frac{1}{M}\sum_{i=0}^{M}scoh_i - \frac{1}{(M(M-1))/2}\sum_{i\neq j}^{M}scop_{i,j}
    \label{eq:sm}
\end{equation}

SM is commonly regarded as an important measure of decomposition quality because it reflects the balance between intra-service cohesion and inter-service coupling~\cite{mitchell2008, filippone2023graphcluster, saied2024migration}. A higher SM value indicates that structurally related classes are assigned to the same service, while unnecessary dependencies between services are minimized. Consequently, SM is frequently used as a primary optimization objective in decomposition approaches.

\subsubsection{Interface Number (IFN)~\cite{fosci2020}}quantifies the average number of interfaces exposed by each microservice and is computed as shown in Equation~\ref{eq:ifn}, where $ifn_i$ is the number of interfaces in the $i^{th}$ microservice. Interfaces refer to the externally accessible entry points through which a service communicates with other services. Lower IFN values indicate simpler, less fragmented services~\cite{saied2024migration}.

\setlength{\abovedisplayskip}{3pt}
\setlength{\belowdisplayskip}{3pt}
\begin{equation}
IFN = \frac{1}{M}\sum_{i=0}^{M}ifn_i
\label{eq:ifn}
\end{equation}

IFN captures service interface complexity; a high IFN may indicate overly fine-grained decomposition or unclear API boundaries. Reducing IFN supports cleaner service design and simpler inter-service communication.

\subsubsection{Inter-partition Communication (ICP)~\cite{kalia2020mono2micro}}measures the proportion of runtime calls that occur across partition boundaries and is computed as shown in Equation~\ref{eq:icp}. It captures the extent to which different services interact during execution. Since communication between microservices can directly affect system performance, particularly response time and resource utilization~\cite{niswar2024performance, bolanowski2022efficiency, tapia2020monolithic}, lower ICP values indicate better service separation and reduced runtime coupling.

\setlength{\abovedisplayskip}{3pt}
\setlength{\belowdisplayskip}{3pt}
\begin{equation}
ICP_{i,j} = \frac{\gamma_{i,j}}{\sum_{i^{'}=0}^{M}\sum_{j^{'}=0, j^{'}\neq i^{'}}^{M}\gamma_{i^{'},j^{'}}}
\label{eq:icp}
\end{equation}

A high ICP value suggests frequent cross-service calls, increasing latency and reducing service autonomy. Minimizing ICP supports loosely coupled services with better potential for independent deployment.

\subsubsection{Non-Extreme Distribution (NED)~\cite{saied2024migration}}assesses how evenly service sizes are distributed within a decomposition. It is defined as shown in Equation~\ref{eq:ned}, where lower NED values reflect more balanced microservice size distributions. Originally proposed by~\citet{wu2005comparison}, this metric identifies disproportionate service partitions where a few microservices dominate in size. A microservice is considered non-extreme when it satisfies $5 \le |m_i| \le 20$~\cite{kalia2021mono2micro,scanniello2010architectural}.

\setlength{\abovedisplayskip}{3pt}
\setlength{\belowdisplayskip}{3pt}
\begin{equation}
NED = \frac{\mathlarger{\sum_{i=0}^{M}} n_i \begin{cases}
0 & 5 \le |m_i| \le 20 \\
1 & Otherwise
\end{cases}}{M}
\label{eq:ned}
\end{equation}

NED evaluates size balance among microservices. Extremely uneven service distributions, where a few services dominate, can lead to scalability bottlenecks. Hence, maintaining moderate NED values ensures that services remain evenly balanced and maintain granularity consistent with the ``small and independent'' nature of microservices~\cite{dragoni2017microservices}.

\subsubsection{Composite Score}

To summarize the overall decomposition quality across multiple metrics, we compute a composite score for each method and benchmark using weighted z-score normalization adapted from \citet{sellami2025contrastive}. For each metric $m \in \{SM, IFN, ICP, NED\}$ and method $t$, the raw metric value $x_{m,t}$ is standardized across all compared methods on the same benchmark. Let $\mu_m$ and $\sigma_m$ denote the mean and standard deviation of metric $m$ over all methods for a given benchmark. The normalized value is computed as:

\begin{equation}
Z_{m,t} = \frac{x_{m,t} - \mu_m}{\sigma_m}
\end{equation}

The composite score for method $t$ is then computed as:

\begin{equation}
Score(t) =
\frac{
\sum_{m \in \{SM,IFN,ICP,NED\}} w_m Z_{m,t}
}{
\sum_{m \in \{SM,IFN,ICP,NED\}} |w_m|
}
\end{equation}

where $w_m$ denotes the weight assigned to metric $m$. Following the weighting convention of \citet{sellami2025contrastive}, metrics that should be maximized are assigned positive weights, while metrics that should be minimized are assigned negative weights. Since SM should be increased, it is assigned a positive weight. IFN, ICP, and NED should be decreased, so they are assigned negative weights.
Following the weighting scheme established in the study
\citep{weerasinghe2026monolith}, we use:

\begin{equation}
W = \{w_{SM}, w_{IFN}, w_{ICP}, w_{NED}\} = \{3, -1, -1, -1\}
\end{equation}

We assign a greater positive weight to SM because it captures the core structural objective of microservice decomposition. However, this weighting makes the composite score sensitive to SM improvements and should be interpreted alongside the individual metrics.

\section{Results}

The following research questions guide the evaluation:

\textbf{RQ1:} How does TripleBound compare with existing structural and semantic decomposition baselines in terms of decomposition quality?

\textbf{RQ2:} How does TripleBound perform across benchmark
systems of different sizes and decomposition granularities?

\textbf{RQ3:} How sensitive is TripleBound to key training parameters such as the hybrid-balance parameter $\alpha$ and the number of training epochs?

\subsection{RQ1: Comparison with Structural and Semantic Baselines}

We compare TripleBound against two representative baselines: CHGNN, a structure-based heterogeneous graph neural network approach, and MonoEmbed, a semantic embedding-based decomposition approach. The experimental results in Table~\ref{tab:decomposition_quality} show that TripleBound achieves the best overall composite score on AcmeAir, DayTrader and JPetStore, while CHGNN performs best on PlantsByWebSphere. However, the picture at the individual metric level is nuanced: the composite score triple-weights SM, and gains on SM and ICP are sometimes accompanied by regressions on NED or IFN. Thus, the composite score is used only as a summary indicator, not as definitive evidence that one decomposition is preferable under all quality criteria.

For DayTrader, TripleBound obtains the highest composite score of 0.7708. It achieves the best SM value of 0.14 and the lowest ICP value of 0.48, indicating improved service cohesion and reduced inter-service coupling. The IFN value of 5.10 is higher than MonoEmbed's best of 1.35, suggesting that the hybrid decomposition exposes more inter-service interfaces on this system. NED also regresses relative to CHGNN (0.62 vs.\ 0.50), indicating a somewhat less balanced entity distribution. The SM gain over CHGNN is small in absolute terms (0.14 vs.\ 0.13), and the practical significance of this difference should be interpreted with caution given the absence of statistical significance testing.

For AcmeAir, TripleBound achieves the highest SM value of 0.23, the lowest IFN value of 2.36, and the lowest ICP value of 0.28, resulting in the highest composite score of 0.8284. A notable limitation is the NED value of 0.71, which is substantially higher than CHGNN's 0.00, indicating that the resulting service sizes are less evenly distributed. This trade-off suggests that improving cohesion and coupling on AcmeAir comes at the cost of entity distribution balance.

For JPetStore, TripleBound achieves the highest composite score of 0.2455. Compared with CHGNN, TripleBound slightly improves IFN (2.95 vs.\ 3.00) and SM (0.18 vs.\ 0.15), while CHGNN obtains the best NED value of 0.00. MonoEmbed achieves the lowest IFN and ICP values, but its lower SM value of 0.03 and higher NED value of 0.72 reduce its overall composite score.

For PlantsByWebSphere, CHGNN achieves the best composite score of 0.5188, outperforming TripleBound on SM, IFN, ICP, and NED. TripleBound's NED of 0.21 is close to CHGNN's 0.20, but its ICP of 0.58 and IFN of 4.31 are worse, indicating that the complete TripleBound configuration does not generalise as effectively to this system. The result establishes an application-specific weakness, although the present evaluation cannot determine which modified component causes it.

\subsubsection{Robustness to Alternative Composite Weights}

To assess whether the aggregate ranking is determined by the threefold weight assigned to SM, we recomputed the composite score from the metric values in Table~\ref{tab:decomposition_quality} using two alternative weight vectors: $W=\{2,-1,-1,-1\}$ and the equal-magnitude vector $W=\{1,-1,-1,-1\}$. This calculation changes only the aggregation of the reported metrics and does not require retraining. Table~\ref{tab:weight_sensitivity} reports the highest-ranked method for each dataset under each weighting.

\begin{table}[t]
\centering
\caption{Highest-ranked method under alternative composite weights. Parentheses contain the corresponding composite score.}
\label{tab:weight_sensitivity}
\resizebox{\linewidth}{!}{
\begin{tabular}{lccc}
\toprule
\textbf{Dataset} & \textbf{$w_{SM}=3$} & \textbf{$w_{SM}=2$} & \textbf{$w_{SM}=1$} \\
\midrule
DayTrader & TripleBound (0.7708) & TripleBound (0.6421) & TripleBound (0.4491) \\
PlantsByWebSphere & CHGNN (0.5188) & CHGNN (0.4088) & CHGNN (0.2437) \\
AcmeAir & TripleBound (0.8284) & TripleBound (0.7421) & TripleBound (0.6126) \\
JPetStore & TripleBound (0.2455) & CHGNN (0.1530) & CHGNN (0.0756) \\
\bottomrule
\end{tabular}
}
\end{table}

The ranking is stable for DayTrader, PlantsByWebSphere, and AcmeAir across these weightings. TripleBound remains first on DayTrader and AcmeAir, whereas CHGNN remains first on PlantsByWebSphere. In contrast, TripleBound's first-place result on JPetStore depends on assigning SM the threefold weight: CHGNN ranks first when the SM weight is reduced to either two or one. Thus, the aggregate evidence supports a weighting-robust advantage for TripleBound on two of the four systems, rather than a weighting-independent advantage on three systems. Raw metrics remain necessary when decomposition priorities differ.

\subsection{RQ2: Performance Across Benchmark Characteristics}

TripleBound exhibits different decomposition trade-offs across the four evaluated systems. To make these differences concrete, we consider the observable characteristics reported in Table~\ref{tab:dataset_stats}, particularly system size and the number of inferred service groups. DayTrader is the largest evaluated system, with 111 ICU classes and 210 service/endpoints entries, and is decomposed into six clusters, whereas AcmeAir and JPetStore are substantially smaller and are each decomposed into four clusters. PlantsByWebSphere represents an intermediate case with five inferred groups.

On the larger DayTrader system, TripleBound improves SM and ICP over both baselines, achieving an SM of 0.14 and ICP of 0.48, although this is accompanied by worse NED than CHGNN and substantially higher IFN than MonoEmbed. On AcmeAir, TripleBound achieves the strongest SM, IFN, and ICP values, but produces substantially worse NED. On JPetStore, the improvement is more limited, with TripleBound improving SM over CHGNN while failing to improve ICP or NED. On PlantsByWebSphere, TripleBound is outperformed by CHGNN across all four metrics.

These results show that TripleBound's gains are not associated uniformly with either larger or smaller systems, nor with a particular number of target clusters. Instead, its effect is metric- and dataset-specific. Because the present evaluation does not directly quantify properties such as dependency density, package modularity, naming consistency, or agreement between parser-inferred groups and structural dependencies, we cannot attribute the observed differences to specific codebase characteristics.

\subsection{RQ3: Sensitivity to Training Parameters}

To answer RQ3, we analyze the sensitivity of TripleBound to the hybrid balance parameter $\alpha$ and the number of training epochs. Each setting was repeated for 30 random initializations, and the reported composite scores correspond to average values across these runs.

First, we varied $\alpha$ from 0.1 to 0.9 while fixing the number of training epochs to 100. Since $\alpha$ controls the balance between local distance preservation and triplet-based semantic supervision, this experiment evaluates how different weighting choices affect decomposition quality.

\begin{figure}[t]
\centering
\includegraphics[width=\columnwidth]{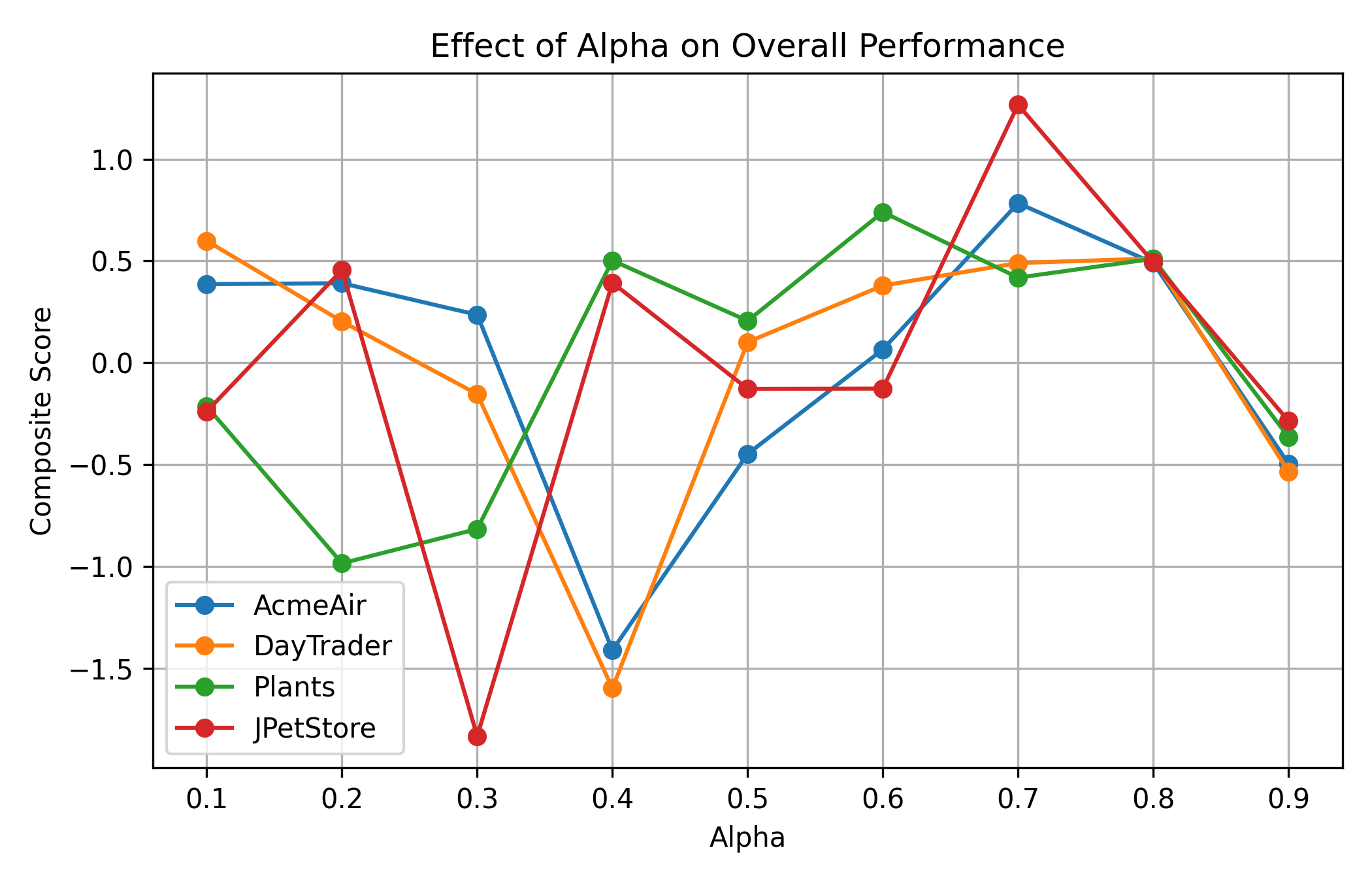}
\caption{Effect of $\alpha$ on overall decomposition performance. Results are averaged across 30 runs with the number of training epochs fixed at 100.}
\label{fig:alpha_composite}
\end{figure}

As shown in Fig.~\ref{fig:alpha_composite}, the effect of $\alpha$ varies across the benchmark systems, indicating that the optimal balance between structural and semantic signals depends on the characteristics of the target application. However, $\alpha = 0.7$ provides a stable trade-off across the evaluated datasets, achieving strong composite scores without the sharp performance drops observed at some lower values.

Next, we varied the number of training epochs from 50 to 250 while fixing $\alpha = 0.7$. This experiment evaluates whether longer training improves the hybrid objective or causes overfitting to dataset-specific structural and organizational patterns.

\begin{figure}[t]
\centering
\includegraphics[width=\columnwidth]{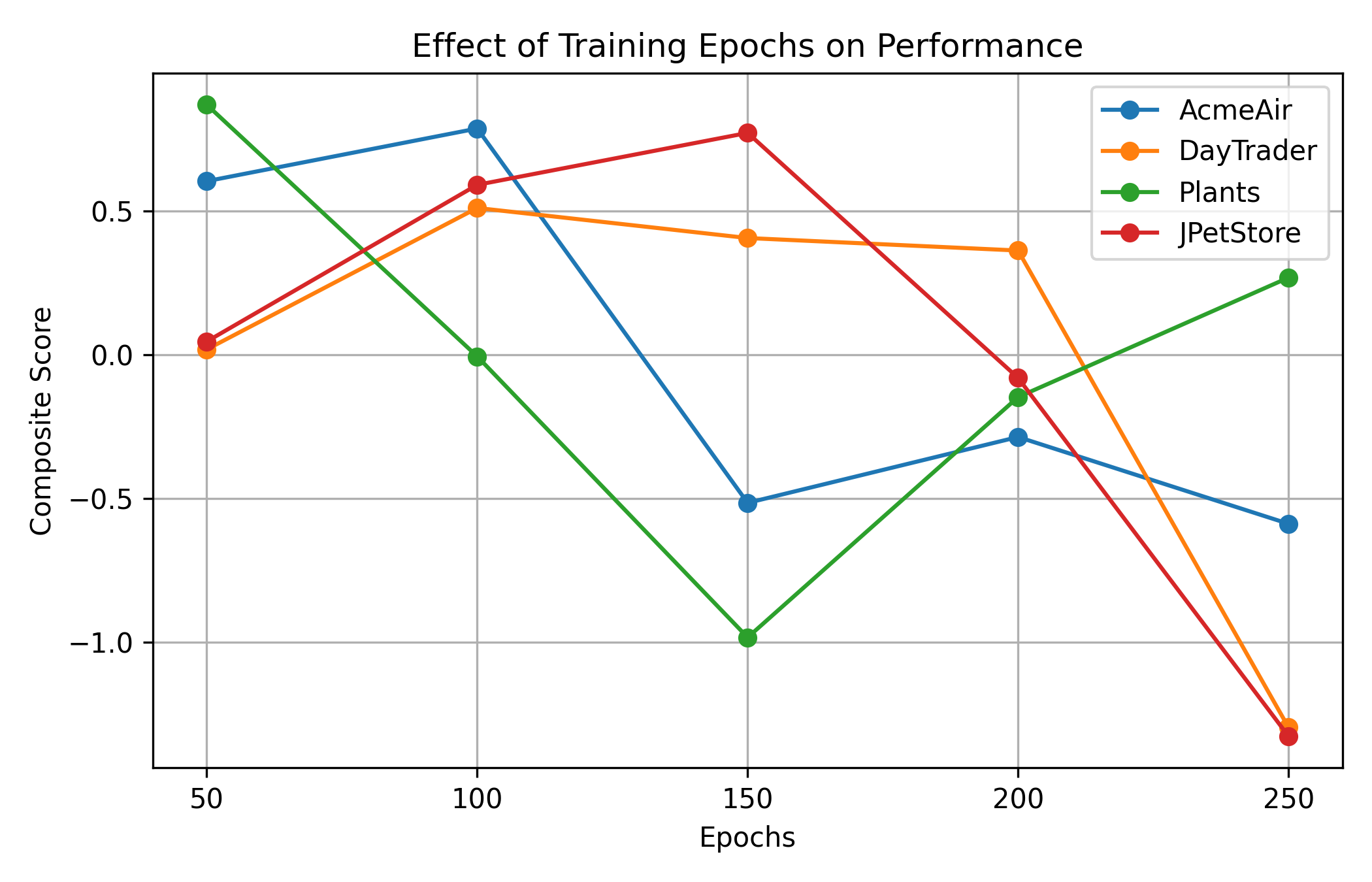}
\caption{Effect of training epochs on overall decomposition performance. Results are averaged across 30 runs with $\alpha = 0.7$.}
\label{fig:epoch_composite}
\end{figure}

As shown in Fig.~\ref{fig:epoch_composite}, 100 epochs provides the best overall balance. Performance tends to degrade at higher values, suggesting that longer training may overfit dataset-specific patterns. Therefore, the final evaluation uses $\alpha = 0.7$ and 100 training epochs.


\section{Discussion}

\subsection{Structural vs.\ Semantic Signal Strength}

The experimental results show that the behavior of the complete TripleBound configuration is application-dependent. On AcmeAir, TripleBound achieves the strongest SM, IFN, and ICP values among the compared methods, but its NED increases substantially from CHGNN's 0.00 to 0.71. Thus, it produces more cohesive and less communication-intensive partitions under the reported metrics at the cost of markedly less balanced service sizes. Whether this trade-off is preferable depends on the architectural priorities of the intended migration. AcmeAir is relatively compact and its package structure
may provide organizational cues that agree with the structural
graph, although this was not directly tested.

On DayTrader, the improvements are more selective. TripleBound improves structural modularity and inter-partition communication. However, this comes with a larger interface surface compared with MonoEmbed and a less balanced entity distribution compared with CHGNN. The complete configuration therefore favors cohesion and coupling on this system but does not uniformly improve interface simplicity or size balance; the contribution of each loss component cannot be inferred from this comparison.

For JPetStore, TripleBound achieves the highest composite score under the selected weighting and improves structural modularity over CHGNN. However, MonoEmbed achieves lower inter-partition communication, CHGNN maintains better entity distribution balance, and the alternative-weight analysis ranks CHGNN first when the SM weight is reduced. The claimed aggregate advantage on this system is therefore not robust to reasonable changes in metric priorities.

\subsection{Failure Mode Analysis: PlantsByWebSphere}

The underperformance of TripleBound relative to CHGNN on PlantsByWebSphere warrants detailed analysis. PlantsByWebSphere is a compact benchmark system, comprising only 36 ICU classes and 47 service endpoints. In such a compact system, the structural dependency graph alone may carry sufficient information for the clustering algorithm to find cohesive service boundaries without additional semantic guidance. Alternatively, triplet constraints derived from package and naming cues may conflict with the structural connectivity patterns. These explanations are plausible but were not directly tested; the available evidence establishes the failure case, not its cause.

This observation is consistent with findings reported in the hybrid decomposition literature~\cite{hierdecomp2019, sellami2022hierarchical}, which suggest that the benefit of combining complementary signals depends critically on the quality and mutual consistency of each individual source. When one signal is already highly informative and well aligned with the desired decomposition outcome, the addition of a weaker or less consistent signal can introduce noise rather than useful supervision. Future work should therefore investigate lightweight signal-quality assessment strategies that can determine, prior to training, whether semantic supervision is likely to improve or degrade decomposition quality for a given target application.

\subsection{Implications for Practice}

These findings carry several practical implications for teams applying automated decomposition tools in real-world modernization projects. First, no single decomposition method consistently dominates across all applications and evaluation metrics. Practitioners should consider the size, organizational structure, and naming conventions of their target system when selecting or configuring a decomposition approach. Applications with clear package hierarchies and consistent naming conventions are more likely to benefit from semantic supervision, while structurally simpler or legacy systems with accumulated technical debt may be adequately served by purely structural methods.

Second, the evaluation metric used significantly influences perceived performance. A method that ranks highest on a composite score may rank lower on criteria most relevant to a given project, such as service size balance (NED) or interface simplicity (IFN). Teams should identify which quality criteria matter most before interpreting results.


Third, the triplet-based supervision strategy shows that weak labels derived from source-level organizational cues can guide representation learning without ground-truth decompositions when those cues align with likely service boundaries. Its effectiveness may decrease in legacy codebases where accumulated technical debt causes source-level organization to diverge from the underlying architectural structure.

Finally, the ICP improvements on AcmeAir and DayTrader are consistent with the intended effect of the Inter-Cluster Communication Loss (\(\mathcal{L}_{icc}\)). They do not establish its independent contribution, however, because the comparison changes triplet supervision, loss weighting, and other training components at the same time. The JPetStore and PlantsByWebSphere results further show that lower ICP is not obtained uniformly.

\section{Threats to Validity}

Several factors may affect the validity of the results reported in this study.

\textit{Limited benchmark coverage.} The evaluation is conducted on four benchmark monolithic systems
highlighted by~\citet{weerasinghe2026monolith}, where an expanded comparative evaluation of decomposition approaches across a broader benchmark set, including additional methods and metrics, is provided.
Although these systems are commonly used in microservice decomposition research, they represent relatively small applications and may not fully capture the complexity and diversity of large industrial monoliths. Generalisation to larger or differently structured systems remains to be validated.  

\textit{Triplet label leakage.} The semantic triplets are derived from parser-inferred service groups based on package structure, naming conventions, and code location. On standard enterprise benchmarks, these organizational cues often correlate strongly with modular boundaries. Consequently, the triplet supervision may partially reflect pre-existing module organization rather than independently discovered semantic relationships, making it difficult to separate meaningful semantic learning from boundary leakage through source organization.

\textit{Hyperparameter tuning on evaluation benchmarks.} The hybrid balance parameter $\alpha$ and the number of training epochs were selected based on sensitivity analyses conducted on the evaluation benchmarks. This entanglement of tuning and evaluation may overestimate performance on the reported benchmarks. Evaluating generalization to held-out systems would strengthen confidence in the chosen configuration.

\textit{Lack of component ablations.} TripleBound changes several aspects of the CHGNN baseline simultaneously, including triplet-based supervision, communication-aware regularization, loss weighting, and removal of the original structure loss. The evaluation therefore establishes the behavior of the complete TripleBound configuration but cannot identify the independent contribution of each modification. In particular, changes in SM or ICP cannot be attributed exclusively to the triplet or communication-aware objectives. An incremental evaluation of the CHGNN backbone, the addition of triplet supervision, the addition of communication regularization, and the complete weighted objective is needed to isolate these effects.

\textit{Statistical uncertainty.} Although each configuration was executed 30 times and Table~\ref{tab:decomposition_quality} reports averages, the evaluation does not report dispersion, confidence intervals, effect sizes, or statistical significance tests. Consequently, small mean differences, such as the DayTrader SM values of 0.14 and 0.13, should not be interpreted as reliable evidence of superiority. Per-run measurements and paired statistical tests with effect sizes and correction for multiple comparisons are required to quantify confidence in the reported differences.

\textit{Composite score weighting.} The composite score uses the weight vector $W=\{3,-1,-1,-1\}$, which gives SM greater influence than IFN, ICP, and NED. This choice follows the view that structural modularity is central to decomposition quality, but it can favor methods that optimize cohesion and coupling while degrading service-size balance. The alternative-weight analysis reduces this concern but cannot cover every project-specific preference. In particular, TripleBound's first-place ranking on JPetStore changes when the SM weight is reduced. Composite scores should therefore be interpreted alongside the raw metrics rather than as an absolute ranking.

\textit{Cluster count selection.} The number of clusters $K$ is set to the number of parser-inferred seed groups. This choice may influence the Normalized Entity Distribution (NED) metric and may not always correspond to the ideal number of microservices for a given system. The impact of alternative cluster selection strategies is not assessed.

\textit{Metric coverage.} The evaluation relies on structural metrics (SM, IFN, ICP, and NED). While these are widely used in decomposition research, they do not capture business-domain alignment, maintainability, deployability, or architectural feasibility. Comparison against expert-designed decompositions or practitioner validation would provide a more complete picture of decomposition quality.

\textit{Parser accuracy.} The quality of the decomposition depends on the accuracy of the parser and the extracted dependency graph. Missing or incorrect dependencies, database interactions, or code-level relationships may affect the learned representations and the resulting service boundaries.

\section{Conclusion}

This paper presented TripleBound, a hybrid framework for automated monolith-to-microservices decomposition. It combines heterogeneous graph-based structural learning with weakly supervised triplet constraints derived from parser-inferred service groups. Unlike post-training fusion, TripleBound injects these constraints directly into the shared latent space, jointly optimizing structural and grouping objectives.

Experiments on four benchmarks show that TripleBound achieves the highest composite score on AcmeAir, DayTrader, and JPetStore under the selected weighting scheme. The ranking remains stable under two alternative weightings on AcmeAir and DayTrader, but not on JPetStore, while CHGNN performs better on PlantsByWebSphere under all tested weightings. The results are metric-selective: improvements in modularity and coupling may coincide with reduced entity distribution balance. Because the current evaluation does not isolate the individual loss components or quantify statistical significance, it supports the complete framework as a promising, application-dependent configuration rather than establishing uniform or component-specific superiority.

Future work will focus on runtime and domain-level signals, adaptive weighting, improved clustering, and feedback-based refinement. TripleBound can also be extended toward migration support through API generation, dependency analysis, communication recommendations, and initial code generation.

{\footnotesize
\bibliographystyle{IEEEtranN}
\bibliography{references}
}

\end{document}